\documentclass[aps,prb,reprint,twocolumn,floatfix,superscriptaddress,longbibliography,groupedaddress]{revtex4-2}

\usepackage{standalone}
\usepackage{amsthm}
\usepackage{lipsum}
\usepackage{amsfonts, amsmath, amssymb}
\usepackage{graphicx}
\usepackage{dcolumn}
\usepackage{bm}
\usepackage{dsfont}
\usepackage[normalem]{ulem}
\usepackage{color}
\usepackage[utf8]{inputenc}
\usepackage[colorlinks,citecolor=blue,linkcolor=blue,urlcolor=blue]{hyperref}
\usepackage{tikz}
\usepackage{braket}
\usepackage{physics}
\usepackage{color}
\usepackage{soul}
\usepackage{mathtools}
\usepackage{float}
\usepackage{esvect}
\usepackage{subfigure}
\usepackage{stmaryrd} 
\usepackage{enumerate}
\usepackage{bbold}

\include{physics}

\newcommand{\SM}{Appendix}

\begin{document}

\title{Electron conductance and many-body marker \\ of a cavity-embedded topological 1D chain}
\author{Danh-Phuong Nguyen}
\affiliation{Universit\'{e} Paris Cit\'e, CNRS, Mat\'{e}riaux et Ph\'{e}nom\`{e}nes Quantiques, 75013 Paris, France}
\author{Geva Arwas}
\affiliation{Universit\'{e} Paris Cit\'e, CNRS, Mat\'{e}riaux et Ph\'{e}nom\`{e}nes Quantiques, 75013 Paris, France}
\author{ Cristiano Ciuti}
\affiliation{Universit\'{e} Paris Cit\'e, CNRS, Mat\'{e}riaux et Ph\'{e}nom\`{e}nes Quantiques, 75013 Paris, France}

\begin{abstract}
\noindent 

We investigate many-body topological and transport properties of a one-dimensional Su–Schrieffer–Heeger (SSH) topological chain coupled to the quantum field of a cavity mode. The quantum conductance is determined via Green's function formalism in terms of the light-matter eigenstates calculated via exact diagonalization for a finite number of electrons. We show that the topology of the cavity-embedded many-electron system is described by a generalized electron-photon Zak marker.  We reveal how the quantization of transport is modified by the cavity vacuum fields for a finite-size chain and how it is impacted by electronic disorder. Moreover, we show that electron-photon entanglement produces dramatic differences with respect to the predictions of mean-field theory, which strongly underestimates cavity-modified transport. 
\end{abstract}
\date{\today}
\maketitle
\section{Introduction}
In recent years there has been a growing interest in manipulating materials using cavity vacuum fields, as evidenced by various studies \cite{FornDaz2019, FriskKockum2019, GarciaVidal2021, Schlawin2022}. Various platforms, such as metallic split-ring terahertz electromagnetic resonators \cite{Keller2017, Scalari2012, ParaviciniBagliani2018}, and more recently, hyperbolic van der Waals materials \cite{Ashida2023}, enable ultra-strong light-matter coupling due to their exceptional sub-wavelength photon mode confinement. On the theoretical front, several models have been proposed to study its impact on diverse aspects like superconductivity \cite{Sentef2018, Curtis2019, Schlawin2019, Kozin2024, Gomez-Leon2024}, quantum transport \cite{Hagenmller2017, Hagenmller2018, Arwas2023,Rokaj2023, Winter2023, Macedo2024}, and topology \cite{Dmytruk2022, Li2022, Shaffer2023, Nguyen2023, Bacciconi2024,Lin2023,Dmytruk2024,Dag2023,PerezGonzalez2023,MndezCrdoba2020,MndezCrdoba2023,Yang2024}. On the experimental front, this intricate physics has been demonstrated through investigations of magneto-transport properties in the Shubnikov-de Haas regime \cite{ParaviciniBagliani2018}, modification of the integer quantum Hall effect \cite{Appugliese2022,Kuroyama2023,Kuroyama2024} and enhancement of fractional quantum Hall gaps \cite{Enkner2024}. Additionally, a study has  reported the modification of the critical temperature of a charge density wave transition \cite{Jarc2023} by the coupling to a tunable cavity. 

To address the challenges posed by the cavity quantum electrodynamics (QED) many-body problem, which involves both fermionic and bosonic particles, several approaches have been suggested in the literature. One such method is the adiabatic elimination, as suggested in Refs. \cite{Ciuti2021, Li2022, Arwas2023, Lin2023, Dag2023}. In this approach, electronic states are coupled thanks to photon-mediated processes, resulting in an effective electronic Hamiltonian. However, it requires an energy scale separation between light and matter degrees of freedom, i.e., photon energy must be off-resonant to the relevant electronic transitions. Another viable approximation is the mean-field ansatz, initially proposed in Ref. \cite{Andolina2019} and then later used in Ref. \cite{Guerci_PRL_2020}. Path integral formalism with additional Gaussian fluctuations are also considered in Refs.  \cite{Dmytruk2022, Dmytruk2024, Guo2024}. This technique works in the thermodynamic limit and it implies no entanglement between light and matter, allowing the ground state to be determined self-consistently through effective photon and electron Hamiltonians with renomalised parameters. However, this semi-classical regime overlooks quantum fluctuations and light-matter entanglement, possibly neglecting the potential emergence of interesting physics. For example, when light and matter are highly entangled, novel phenomena such as light-matter Chern numbers \cite{Nguyen2023} and Majorana polaritons \cite{Bacciconi2024} can arise. This regime could be studied through bosonization of fermionic particles \cite{Dmytruk2022,Rokaj2023}, Density Matrix Renormalization Group (DMRG) \cite{MndezCrdoba2020,MndezCrdoba2023, Bacciconi_ladder_2023,Bacciconi2024,Bacciconi2024_B}, or exact diagonalization \cite{Nguyen2023,PerezGonzalez2023}.  
\begin{figure}[t!]
\includegraphics[width = 1.0\hsize]{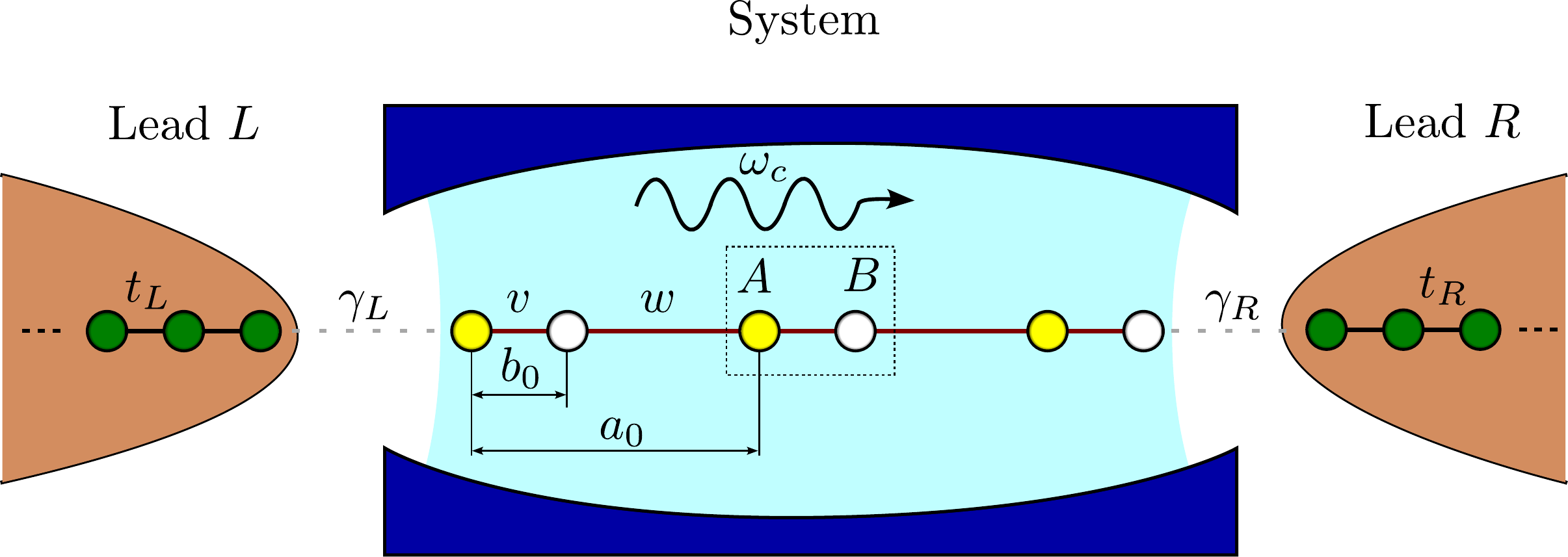}
  \caption{Sketch of a cavity-embedded Su-Schrieffer-Heeger (SSH) chain, consisting of two sublattices $A$ and $B$. The intra-cell (inter-cell) distance is $a_0$ ($b_0$). The corresponding electron hopping coupling is $v$ ($w$). The cavity photon mode frequency is $\omega_c$. The left (right) lead has hopping coefficient $t_L$ ($t_R$), while its coupling to the system is $\gamma_L$ ($\gamma_R$).}
\label{manu_fig:Sys}
\end{figure}

 In this article, we investigate two fundamental problems concerning a paradigmatic cavity-embedded Su-Schrieffer-Heeger (SSH) system \cite{Dmytruk2022,PerezGonzalez2023} by exploiting exact diagonalization for finite-size systems. First, we show the existence of a many-body electron-photon topological marker that generalizes the electronic marker introduced in Refs. \cite{Gilardoni2022, Molignini2023}. Second, we obtain exact results for the electron quantum conductance by a Green's function formalism \cite{Meir1992} by exploiting the exact light-matter eigenstates. We show how the cavity vacuum fields can dramatically affect the transport, as well as a breakdown of mean-field theory due to sizeable electron-photon entanglement. 

\section{Cavity QED Hamiltonian} Let us consider the Hamiltonian describing an SSH chain with $N$ unit cells with two sublattices denoted as $A$ and $B$ (Fig. \ref{manu_fig:Sys}):
\begin{equation}
   \label{manu_eq:SSH} \hat{\mathcal{H}}_{\text{SSH}}= v\sum_{n=1}^{N}\hat{c}^{\dagger}_{n,B}\hat{c}_{n,A} + w\sum_{n=1}^{N-1}\hat{c}^{\dagger}_{n+1,A}\hat{c}_{n,B} + \text{h.c.} \, ,
\end{equation}
where $\hat{c}^{\dagger}_{n,\sigma}$ creates an electron  on the site $n$ and sublattice $\sigma \in \{A,B\}$. Note that we have considered open boundary conditions for the chain. For simplicity, we will consider only one spin channel and omit the spin index. The intra-cell and inter-cell  hopping amplitudes are denoted respectively as $v$ and $w$. When $\vert v \vert > \vert w \vert$ the system is known to be topologically trivial, while $\vert v \vert < \vert w \vert$ gives a non-trivial topology  with localised edge states and non-zero Zak phase at half-filling \cite{Gilardoni2022,Molignini2023}. For the photonic component, we consider a model describing a single-mode cavity with spatially homogeneous field described by the vector potential $\hat{\mathbf{A}} = A_0\mathbf{u}(\hat{a}+\hat{a}^{\dagger})$ where $\hat{a}^{\dagger}$ ($\hat{a}$) represents the photonic creation (annihilation) operator, and $\mathbf{u}$ is the orientation of the cavity mode polarization. For sake of simplicity, we will consider a linear polarization along the chain direction. We can express the vacuum field amplitude as $A_0 = \sqrt{\hbar\omega_c/(2\epsilon_0V_{mode})}$ where $\omega_c$ is the cavity frequency and $V_{mode}$ the mode volume. Light-matter interaction can be introduced via the Peierls substitution giving the following cavity QED Hamiltonian \cite{Dmytruk2022,PerezGonzalez2023}:
\begin{widetext}
\begin{equation}
    \label{manu_eq:H}
    \hat{\mathcal{H}}_S \begin{aligned}
    &= \hbar\omega_c\hat{a}^{\dagger}\hat{a} + \left(ve^{-ig_v\left(\hat{a}+\hat{a}^{\dagger}\right)}\sum_{n=1}^{N}\hat{c}^{\dagger}_{n,B}\hat{c}_{n,A} + we^{-ig_w\left(\hat{a}+\hat{a}^{\dagger}\right)}\sum_{n=1}^{N-1}\hat{c}^{\dagger}_{n+1,A}\hat{c}_{n,B} + \text{h.c.}\right).
    \end{aligned}
\end{equation}
\end{widetext}
Let $a_0$ be the lattice constant and $b_0$ the intra-cell distance such that $0 \leq b_0/a_0 \leq 1$. Then, we can introduce dimensionless light-matter coupling constants $g_{v} = gb_0/a_0$ and $g_w = g(1-b_0/a_0)$ where $g = eA_0a_0/\hbar$. In what follows, we will investigate finite-size chains and we will employ exact diagonalization techniques to determine exact results beyond the single-particle approximation and without adiabatic elimination of the photonic degrees of freedom.

\section{Many-body light-matter topological marker}
Light-matter topological invariants have been recently studied for cavity-embedded 2D systems \cite{Nguyen2023} where electron-photon Chern numbers have been uncovered.  They have been also discussed for the cavity-embedded SSH model in Ref. \cite{PerezGonzalez2023}, in which a winding number (although not exactly quantized) has been computed using Green's function formalism with periodic boundary conditions \cite{Gurarie2011,Essin2011}.  However, for transport problems, the natural boundary conditions are open. Here, we introduces a different topological marker that does not rely on periodic boundary conditions and that is quantized for any arbitrary light-matter coupling.  Such topological invariant is calculated from the many-body ground state of Hamiltonian (\ref{manu_eq:H}). \color{black} For the case of a pure electronic system, a many-body topological $Z_2$ invariant has been introduced in Refs. \cite{Gilardoni2022,Molignini2023} by exploiting the concept of Resta's polarization $\hat{\mathcal{P}}$ \cite{Resta1998}:
\begin{equation}
    \label{manu_eq:P}
    \begin{aligned}
        &\hat{\mathcal{P}} = \text{exp}\left(\frac{i2\pi}{N}\hat{R}\right),\\
        &\hat{R} = \sum_{n=1}^Nn\hat{c}^{\dagger}_{n,A}\hat{c}_{n,A} + \left(n+\frac{1}{2}\right)\hat{c}^{\dagger}_{n,B}\hat{c}_{n,B},
    \end{aligned}
\end{equation}
where $\hat{R}$ is the position operator. For the present 1D many-body system, we have found the following topological light-matter marker:
\begin{equation}
\label{manu_eq:Zep}
    \begin{aligned}
        Z^{(e-p)} &= \frac{2}{\pi}\text{Im }\text{log } \langle GS, N \vert \hat{\mathcal{P}} \otimes \hat{I}_p \vert GS, N \rangle,\\
        &= \frac{2}{\pi}\text{Im }\text{log }\text{Tr}_{el}(\hat{\rho}_{el}\hat{\mathcal{P}}),
    \end{aligned}
\end{equation}
where $\vert GS, N \rangle $ is the exact ground state and $N$ is the number of electrons.  Throughout the paper we have focused on the physics at the half-filling point $N_e = N$.
\begin{figure}[t!]
\includegraphics[width = 1.0\hsize]{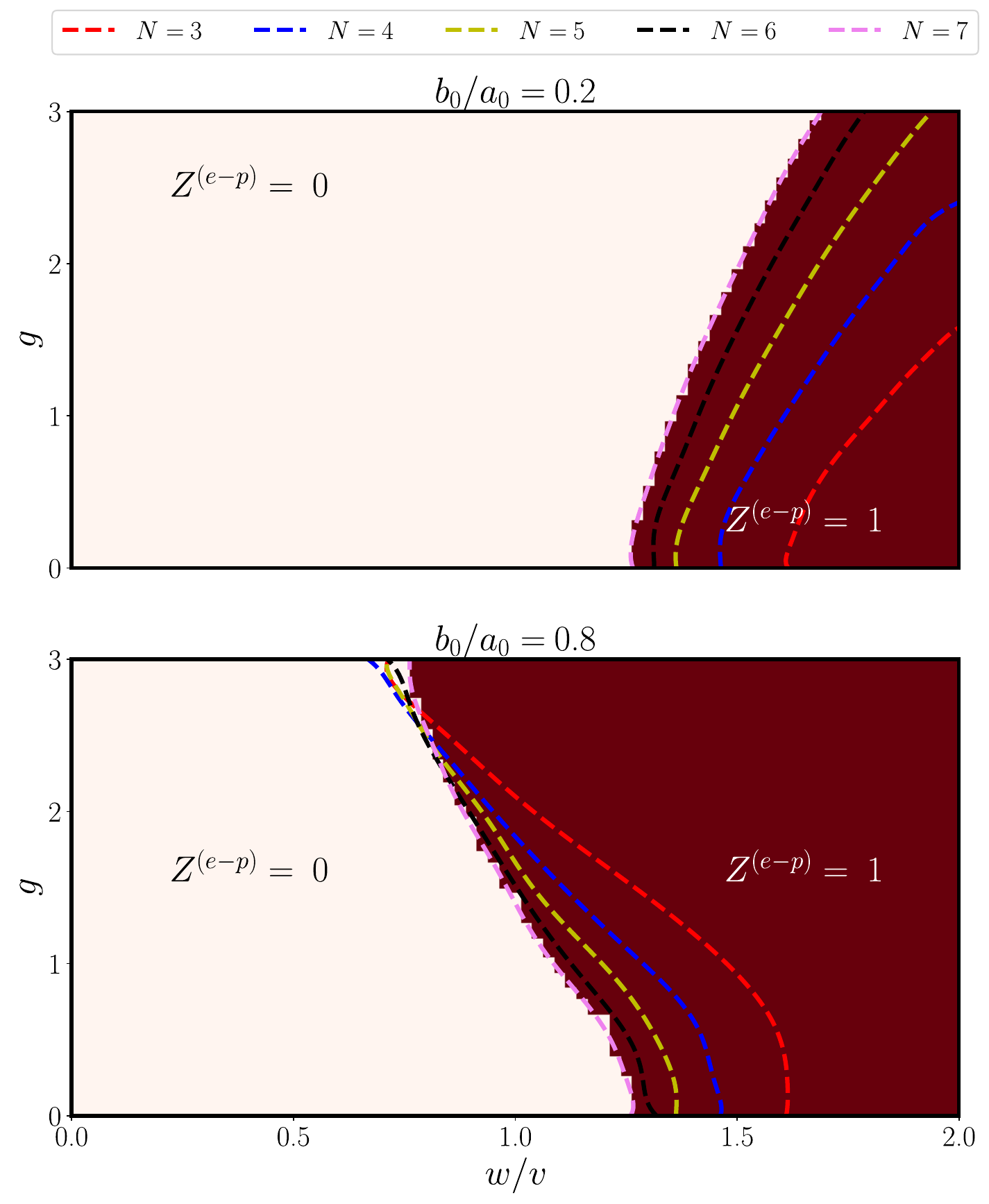}
  \caption{Many-body light-matter topological marker at half-filling, plotted as a function of $w/v$ and $g$ for $b_0 = 0.2a_0$ (top) and $0.8a_0$ (bottom)  for various values of $N$, the number of unit cells. Other parameter: $\hbar\omega_c/v = 1$.
\label{manu_fig:Z}}
\end{figure}
Note that $\hat{I}_p$ is the identity operator for the photon degrees of freedom. The electronic reduced density matrix in the formula above is obtained by tracing over the photonic degrees of freedom, namely $\hat{\rho}_{el}(N) = \text{Tr}_{\text{phot}}\left(\vert GS, N \rangle \langle GS, N \vert\right)$. The second line in Eq. (\ref{manu_eq:Zep}) shows it coincides with the ensemble geometric phase for density matrices \cite{Bardyn2018,Mink2019,Unanyan2020,Wawer2021,Molignini2023}.
Importantly, the reduced electron density matrix $\hat{\rho}_{el}$ has a mixed structure due to light-matter entanglement produced by the cavity coupling.  Note that if you consider a bare electronic system (no cavity coupling) in a mixed thermal state such a marker is no longer integer \cite{Molignini2023} (we have explicitly verified it).  The light-matter entanglement structure is crucial for the marker quantization. As shown in Fig. \ref{manu_fig:Z}, such a quantity is precisely an integer for $N \geq 3$ also for open boundaries. Note that the open boundary conditions produces a small shift with respect to periodic boundary conditions in the topological transition boundary for a finite-size system. 
 
\begin{figure}[t!]
\includegraphics[width = 1.0\hsize]{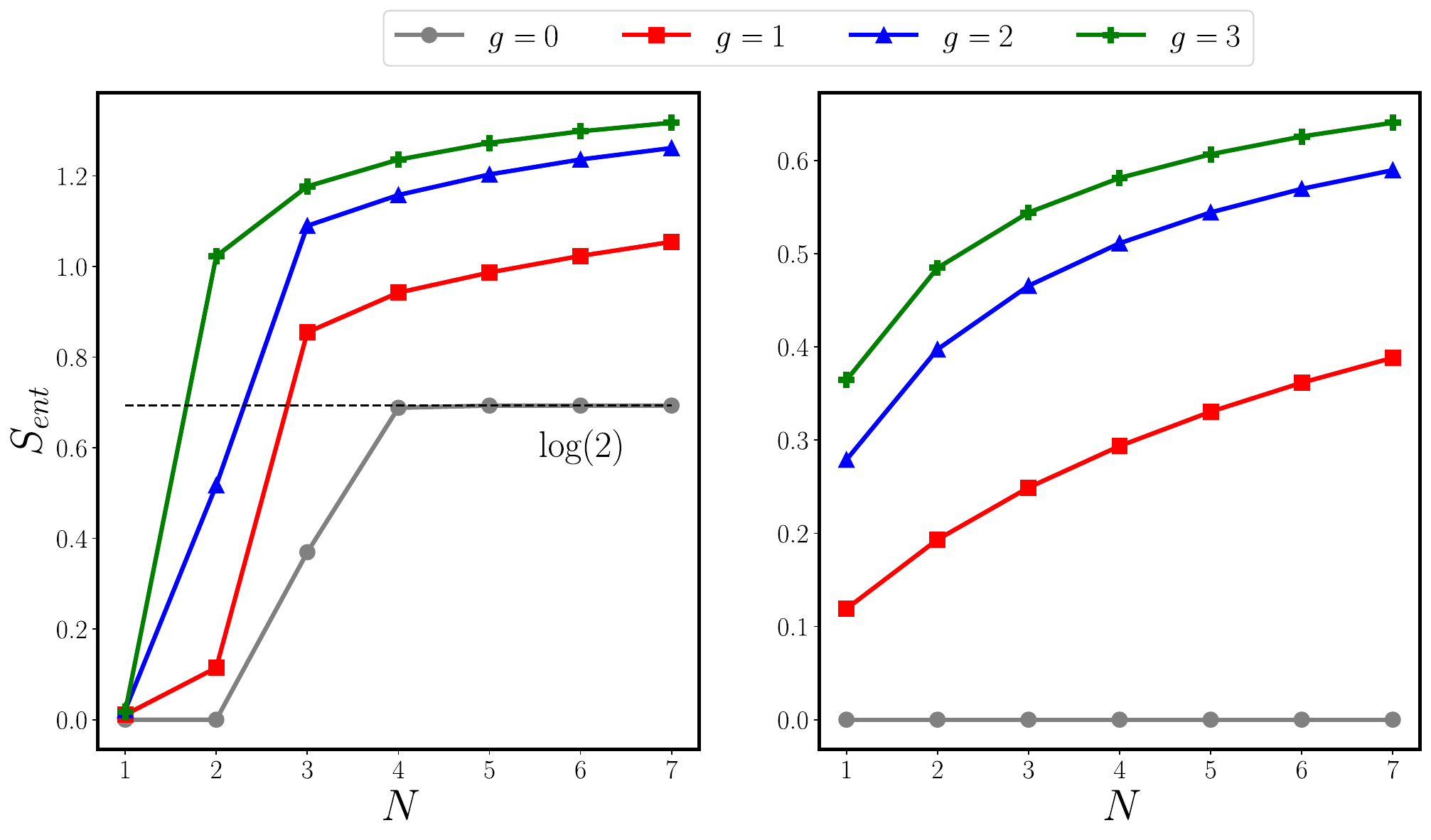}
  \caption{Light-matter entanglement entropy as a function of system size $N$ for different values of the dimensionless light-matter coupling ($g = 0$, $1$, $2$, $3$). Other parameters: $b_0 = 0.5a_0$, (left) $v = 0.1w$, $\hbar\omega_c = v$, (right) $w = 0.1v$, $\hbar\omega_c = w$. Please note that in the left panel where $v/w = 0.1$, the SSH chain is in its degenerate topological ground state, resulting in a $S_{ent}$ value of $\text{log}(2)$.}
\label{manu_fig:PE}
\end{figure}

 Light-matter entanglement can be then quantified by the entanglement entropy $S_{ent} = -\text{Tr}_{\text{el}}(\hat{\rho}_{el}\text{log}\hat{\rho}_{el})$.
In Fig. \ref{manu_fig:PE} we report the entanglement entropy as a function of system size $N$ for different values of the light-matter coupling $g$.
For increasing light-matter coupling, $S_{ent}$ increases and tends to saturate as reported in  Ref. \cite{Shaffer2023}.

\section{Cavity-modified electronic transport} Having the exact light-matter eigenstates, we can study the electron transport properties in the presence of the cavity in the linear response regime. The electronic system $(S)$ is assumed to be  coupled to Left $(L)$ and Right $(R)$ leads (Fig. \ref{manu_fig:Sys}), as described by the Hamiltonian:
\begin{equation}
    \hat{\mathcal{H}} = \hat{\mathcal{H}}_S + \hat{\mathcal{H}}_L +    \hat{\mathcal{H}}_R + \hat{V},
\end{equation}
where $\hat{\mathcal{H}}_S$ has already been shown in Eq. (\ref{manu_eq:H}). The lead Hamiltonians  are   $\hat{\mathcal{H}}_{\lambda} = -t_{\lambda}\sum_{n=1}^{N_{\lambda}}\hat{d}^{\dagger}_{\lambda,n+1}\hat{d}_{\lambda,n} + \text{h.c.}$ with $\lambda \in \{ L, R \}$. Note that we consider the limit of large-size leads, impling that their chemical potentials  $\mu_{\lambda}$ are unaffected by the coupling to the finite-size chain. The coupling between system and leads reads:
\begin{equation}
    \label{manu_eq:V}
    \hat{V} = \left(\gamma_L\hat{c}^{\dagger}_{1,A}\hat{d}_{L,1} + \gamma_R\hat{c}^{\dagger}_{N,B}\hat{d}_{R,1} + \text{h.c.}\right), 
\end{equation}
where $\gamma_{\lambda}$ are the coupling constants.

To determine the quantum transport, we have generalized the framework for interacting conductors derived by Meir and Wingreen \cite{Meir1992} to account for quantum light-matter interaction. The current $J$ flowing through the system in the steady state depends on the Green's functions through the formula:
\begin{equation}
\label{manu_eq:J}
    \begin{aligned}
        J = \frac{ie}{2h}\int d\epsilon &\text{Tr}\{\left[f_L(\epsilon)\Gamma^L - f_R(\epsilon)\Gamma^R\right]\left(G^r - G^a\right)\}\\ 
        + &\text{Tr}\{\left(\Gamma^L - \Gamma^R\right)G^<\},
    \end{aligned}
\end{equation}
where $f_{\lambda}$ represent Fermi-Dirac distribution of the lead $\lambda$ with chemical potential $\mu_{\lambda}$. The matrices $\Gamma^{\lambda}$ depend on the retarded and advanced self-energy terms due to coupling to the corresponding lead, namely $\Gamma^{\lambda} = i(\Sigma^r_{\lambda} - \Sigma^a_{\lambda})$. The propagators $G^r$, $G^a$, $G^<$ are respectively the non-equilibrium retarded, advanced and lesser electronic Green functions of the system in the presence of both leads and light-matter interaction.  It is important to note that the matrices involved in (\ref{manu_eq:J}) have a dimension given  by the system's electronic single-body Hilbert space. A detailed derivation of the Green functions and self-energies can be found in \cite{Meir1992, Datta1995,Arwas2023} and our \SM.

The non-equilibrium Green's functions satisfy the following  equations \cite{Datta1995,Thygesen2008,Arseev2015}:
\begin{equation}
    \begin{aligned}
        &G^r = g_0^r + g_0^r\left(\Sigma^r_{\text{leads}} + \Sigma^r_{\text{int}}\right)G^r,\\
        &G^a = [G^r]^{\dagger},\\
        &G^< = G^r\left(\Sigma^<_{\text{leads}} + \Sigma^<_{\text{int}}\right)G^a.
    \end{aligned}
\end{equation}
\begin{figure}[b!]
\includegraphics[width = 1.0\hsize]{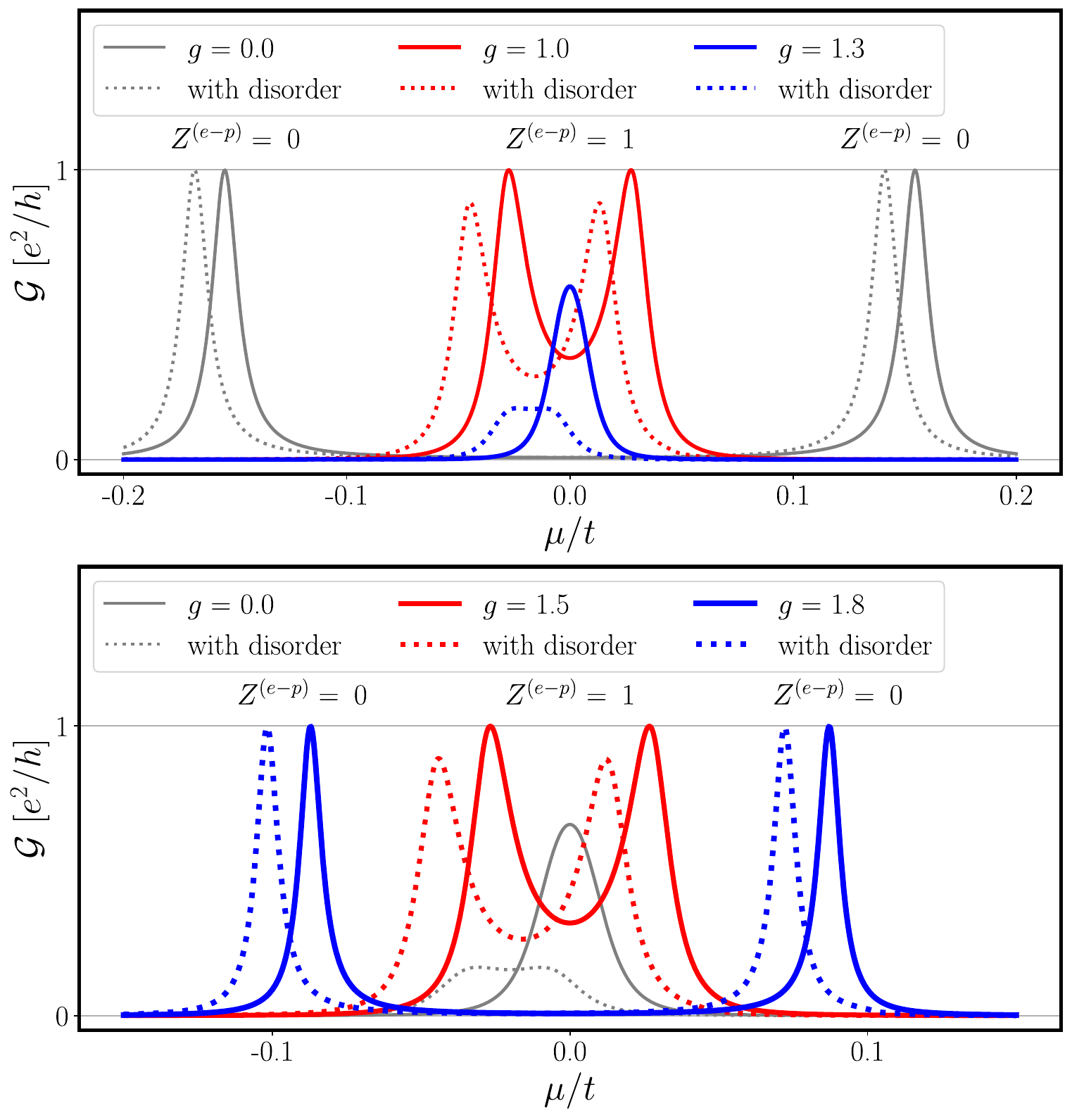}
\caption{Electron conductance $\mathcal{G}(\mu)$ calculated near the half-filling point ($\mu = 0$) versus $\mu/t$ with different values of the dimensionless coupling strength $g$ for two scenarios (top and bottom panels) without (solid lines) or with (dotted lines) random electronic disorder with energy amplitude $W = 0.05 t$. The value of the topological marker $Z^{(e-p)}$ for the peaks of conductance is indicated in the plot. Top: increasing the cavity coupling $g$ reduces the gap between quantized conductance peaks and the quantization is eventually lost. Parameters: $v = 1.0t$, $w = 1.2t$, $b_0 = 0.8a_0$. Bottom: increasing the cavity coupling opens a gap and the conductance becomes quantized. Parameters $v = 0.6t$, $w = 1.4t$, $b_0 = 0.1a_0$. Common parameters for the two panels: $N = 6$, $\gamma_L = \gamma_R = 0.13t$, $\hbar\omega_c = 5.0t$, $t_L = t_R = t$.}
\label{manu_fig:G}
\end{figure}
Here $g_0^{r}(\omega)$ denotes the electron's retarded Green functions of the system without light-matter interaction and without coupling to the leads. The retarded and lesser self-energies due to leads $\Sigma^{r,<}_{\text{leads}}(\omega) = \Sigma^{r,<}_L(\omega) + \Sigma^{r,<}_R(\omega)$ are provided in the \SM, while $\Sigma^{r,<}_{\text{int}}$ arises from light-matter interaction. The latter in general depends on the number of electrons within the system \textit{after} coupling to the leads, therefore on $\mu_{L}$ and $\mu_R$. This term is often treated perturbatively and self-consistently, as discussed in references \cite{Datta1995,Thygesen2008}. Such dependence can be neglected in the weak tunneling limit between the systems and leads, i.e., $\gamma_{L,R}/t \ll N$ in Eq. (\ref{manu_eq:V}). Under this approximation, $\Sigma_{\text{int}}^{r,<}$ can be derived exactly from (\ref{manu_eq:H}). This procedure generates results that are non-perturbative with respect to light-matter interaction, but perturbative in the coupling to the leads. Detailed calculations are given in \SM. 

In presence of a voltage bias $V$ such that $\mu_{L,R} = \mu \pm V/2$, a current flows through the system. At zero temperature the linear conductance reads:
\begin{equation}
    \mathcal{G} = \lim_{V\to 0} \frac{J}{V} = 
    \label{manu_eq:G}
    \frac{e^2}{h}\text{Tr}\left[\frac{i}{4}\Gamma^p\left(G^r -G^a\right)-\frac{1}{4}\Gamma^m G^r \Gamma^m G^a\right],
\end{equation}
where $\Gamma^{p,m} = \Gamma^{L} \pm \Gamma^{R}$. Note that we have omitted the dependence on the chemical potential in the mathematical notation for the sake of simplicity. The results are illustrated in Fig. \ref{manu_fig:G} where two different scenarios ($g_v > g_w$ versus $g_v < g_w$) are reported in two corresponding panels (top and bottom, respectively). In the top panel, we consider the chain to be in a topologically trivial phase with $Z^{(e-p)} = 0$ in the absence of light-matter interaction ($g = 0$).  In such a configuration, the states are delocalized over the finite-size chain and are equally coupled to the two leads due to the inversion symmetry of Hamiltonians (\ref{manu_eq:SSH}) and (\ref{manu_eq:H}), leading to the observation of a quantized conductance $e^2/h$ when the system chemical potential $\mu$ is properly aligned.  Due to the electron-hole symmetry, two identical peaks are symmetrically located around the half-filling point. Increasing the cavity coupling reduces the gap between the two peaks, eventually producing a topological transition towards the non-trivial phase with $Z^{(e-p)} = 1$. Increasing enough $g$, the topological edge states becomes more and more localized producing an eventual loss of the conductance quantization.  Due to the finite size of the considered transport system, the conductance peaks arise mainly from the topological edge states (see Appendix for more details).  Note that the dotted curves correspond to the electron conductance in the presence of disorder. We have considered here an on-site electronic energy random disorder with an uniform probability distribution in the interval $[-W,W]$. In the figure, we have considered $W = 0.05t$. Since $2W$ is smaller than the energy gap between the conductance peaks for the $g=0$ case, the effect of disorder is to produce a shift of the conductance peak energy with $Z^{(e-p)} = 0$, but its quantization holds. Instead, $2W$ is larger than the energy splitting for $g=1$ and $g=1.3$, resulting in a loss of quantization for $g=1$ and a dramatic modification of the peak for $g=1.3$. Note that we have calculated a large number of disorder random realizations (not shown) and the same qualitative picture is consistently observed, apart from quantitative variations. 

The opposite scenario is achieved in the bottom panel where $g_v < g_w$.
Here for no cavity coupling ($g = 0$), the SSH chain is topologically non-trivial with a non-quantized single conductance peak. Increasing the cavity coupling opens a gap and the conductance become quantized at the peaks. 
The conductance with disorder (dotted lines) shows a similar behavior as in the top panel. In the \SM, we have reported other examples of results for larger and smaller values of the disorder energy amplitude $W$.

Note that our theory can be also applied to the case of a spatially inhomogeneous cavity mode field. As an illustrative example, we have considered a mode with a finite and constant spatial gradient in  the \SM. In this scenario, the eigenstates of Hamiltonian (\ref{manu_eq:H}) loses inversion symmetry and become asymmetrically coupled to the leads, resulting in a non-quantized conductance $\mathcal{G}$ in the trivial phase. 

\begin{figure}[t!]
\includegraphics[width = 1.0\hsize]{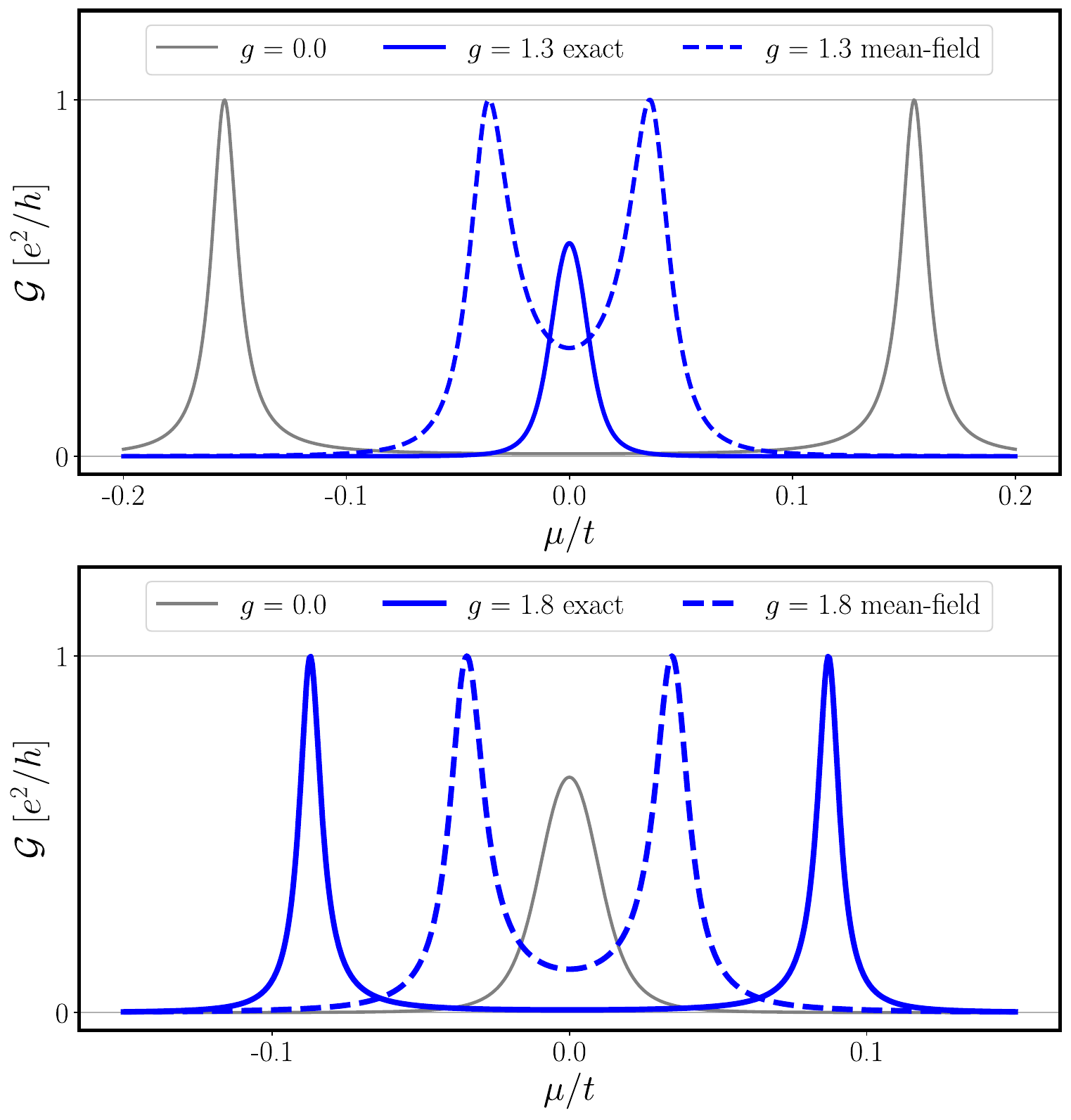}
  \caption{Comparison between conductances calculated from exact diagonalization (solid line) and mean-field (dashed line). Same parameters as in Fig. \ref{manu_fig:G}.
\label{manu_fig:compare}}
\end{figure}
In Fig. \ref{manu_fig:compare} we compare the conductance calculated with exact diagonalization to the predictions of  mean-field theory, where the ground state of (\ref{manu_eq:H}) is approximated as separable (no electron-photon entanglement), namely $\vert GS^{(e-p)} \rangle \simeq \vert \psi^{(e)}\rangle \vert\chi^{(p)}\rangle$. The latter is equivalent to consider a purely electronic model with hopping parameters $v$ and $w$ that are self-consistently renormalized by the photon field  (for details see Supplementary Material). Our results show a breakdown of mean-field theory, which is due to the significant degree of light-matter entanglement shown in Fig. \ref{manu_fig:PE}. In all scenarios, the entanglement significantly enhances the impact of the cavity quantum field on the electronic conductance.\\

\section{Conclusions}
In this Letter, we provided the first study of cavity-modified topology and transport beyond the single-particle picture and without adiabatic elimination of the photon degrees of freedom by exploiting an exact diagonalization approach. We focused our study on a paradigmatic topological 1D system, namely an SSH chain. We have introduced a novel topological many-body marker for the light-matter system, which is a generalization of the Zak phase, but that is valid for any arbitrary light-matter coupling and entanglement. Moreover, we have studied cavity-modified electron transport in the linear regime for finite-size chains by exploiting the many-body eigenstates, showing the crucial role of light-matter entanglement.

\acknowledgements
{We acknowledges financial support from the French agency ANR through the project CaVdW (ANR-21-CE30-0056-0) and from the Israeli Council for Higher Education - VATAT.}


\setcounter{equation}{0}
\renewcommand{\theequation}{A\arabic{equation}}
\setcounter{section}{0}
\renewcommand{\thesection}{APPENDIX \Alph{section}:}

\makeatletter
\renewcommand{\@seccntformat}[1]{\thesection\hspace{1em}}  
\makeatother


\section{Green's functions}
\label{SM:GF}
While employing the non-equilibrium Green's function formalism, we will use the symbols '$r$' and '$a$' to denote retarded and advanced Green's functions respectively.
Moreover '$<$' and '$>$' will refer to lesser and greater Green's functions respectively. We will consider four different cases depending on the presence (or not) of leads and of a cavity. In particular, we will adopt the following notations for the Green's functions of the system only: $g_0$ for the case without leads, without cavity; $g$ without leads, with cavity; $G_0$ with leads, without cavity; $G$ with leads, with cavity. While describing both system and leads, the Green functions will be denoted as $\tilde{g}_0, \tilde{g}, \tilde{G}_0$ and $\tilde{G}$.

The system Hamiltonian $\hat{\mathcal{H}}_S$ described in the main manuscript conserves the number $N_e$ of fermionic particles. Therefore, the exact eigenstates have the form $\vert \chi, N_e \rangle$ and energies $\epsilon_{\chi,N_e}$ where $\chi$ labels eigenstates in the $N_e$-electron eigenspace ${\mathcal S}_{N_e}$. 
The retarded Green's function matrix can be expressed in terms of such many-body eigenstates as follows:
\vspace{0.8cm}
\begin{widetext}
\begin{equation}
    \label{SM_eq:int-Gr}
    \begin{aligned}
    g^r_{ab}(\omega, N_e) &= \sum_{\xi \in {\mathcal S}_{N_e+1}}\frac{\langle GS, N_e \vert \hat{c}_a \vert \xi, N_e + 1\rangle \langle \xi, N_e + 1 \vert \hat{c}_b^{\dagger}\vert GS, N_e \rangle}{\omega - \omega_{\xi,N_e+1} + \omega_{GS,N_e} + i\eta}
   + \sum_{\xi \in  {\mathcal S}_{N_e-1}}\frac{\langle GS, N_e \vert \hat{c}_b^{\dagger} \vert \xi, N_e - 1\rangle \langle \xi, N_e - 1 \vert \hat{c}_a\vert GS, N_e \rangle}{\omega - \omega_{GS,N_e} + \omega_{\xi,N_e-1} + i\eta}.
    \end{aligned}
\end{equation}
\end{widetext}
The expression (\ref{SM_eq:int-Gr}) is known as Lehmann representation. If we denote $g^r_0(\omega,N_e) = g^r_0(\omega)$ as the retarded Green function for the non-interacting case, then the retarded self-energy term $\Sigma_{\text{int}}^{r}(\omega,N_e)$ is defined as:
\begin{equation}
    \label{SM_eq:int-Sigmar}
    \Sigma^r_{\text{int}}(\omega, N_e) = [g^{r}_0(\omega)]^{-1} - [g^{r}(\omega,N_e)]^{-1}.
\end{equation}

In order to describe the effect of two leads on the system, we consider the Hamiltonian:
\begin{equation}
\label{SM_eq:Hall}
    \begin{aligned}
        \hat{\mathcal{H}} &= \hat{\mathcal{H}}_S + \hat{\mathcal{H}}_{\text{leads}} +\hat{V},
    \end{aligned}
\end{equation}
with $\hat{\mathcal{H}}_{\text{leads}} = \sum_{n, \lambda = L, R}t_{\lambda}\hat{d}^{\dagger}_{\lambda,n+1}\hat{d}_{\lambda,n}$ and the tunneling term $\hat{V} = \hat{v} + \hat{v}^{\dagger}$ with  $\hat{v} = \gamma_L\hat{c}^{\dagger}_{1,A}\hat{d}_{L,1} + \gamma_R\hat{c}^{\dagger}_{N,B}\hat{d}_{R,1}$. In the non-interacting case, the single-body retarded Green function for the system and leads can be employed to simplify the problem: $\tilde{G}_0^r(\omega) = (\omega - \hat{\mathcal{H}} + i\eta)^{-1}$. More specifically, its block form is expressed as:
\begin{equation}
\label{SM_eq:Grleads}
\begin{aligned}
    \tilde{G}_0^r(\omega) &= \begin{pmatrix}
    \omega - \hat{\mathcal{H}}_S + i\eta && -\hat{v} \\ -\hat{v}^{\dagger} && \omega - \hat{\mathcal{H}}_{\text{leads}} + i\eta \end{pmatrix}^{-1}\\
    &= \begin{pmatrix} 
    [g_{0,S}^r]^{-1}(\omega) && -\hat{v} \\ -\hat{v}^{\dagger} && [g_{0,\text{leads}}^r]^{-1}(\omega)
    \end{pmatrix}^{-1},     
\end{aligned}
\end{equation}
where $g_{0,S}^r$ and $g_{0,\text{leads}}^r$ represent the retarded Green functions of the system and the leads when they are not coupled. The Green function for the system is obtained through block matrix inversion:
\begin{equation}
\begin{aligned}
    G^r_{0,S}(\omega) &= g^r_{0,S}(\omega)\left[1 - \hat{v}g_{0,\text{leads}}^r(\omega)\hat{v}^{\dagger}g^r_{0,S}(\omega)\right]^{-1}\\
    &= g^r_{0,S}(\omega)\left[1 - \Sigma^r_{\text{leads}}(\omega)g^r_{0,S}(\omega)\right]^{-1}.
\end{aligned}
\end{equation}
In this context, the retarded self-energies induced by leads $L, R$ are defined as $\Sigma_{\text{leads}}^r(\omega) = \hat{v}g^r_{0,\text{leads}}(\omega)\hat{v}^{\dagger} = \hat{v}g^r_{0,L}(\omega)\hat{v}^{\dagger} + \hat{v}g^r_{0,R}(\omega)\hat{v}^{\dagger} = \Sigma_{L}^r(\omega) + \Sigma_{R}^r(\omega)$. They read:
\begin{equation}
    \label{SM_eq:SF_leads}
    \begin{aligned}
        &\Sigma_{L}^r(\omega) = -\gamma_{L}^2/t_{L}\text{exp}\left(ik_L\right)\vert 1, A \rangle \langle 1, A \vert,\\
        &\Sigma_{R}^r(\omega) = -\gamma_{R}^2/t_{R}\text{exp}\left(ik_R\right)\vert N, B \rangle \langle N, B \vert,
    \end{aligned}
\end{equation}
with $\omega = -2t_{\lambda}\text{cos}(k_{\lambda})$. The states $\vert 1, A \rangle$ and $\vert N, B \rangle $ are single-body states localized respectively at the beginning and end of the SSH chain. Another relevant quantity is described by the matrix $\Gamma_{\lambda} = i[\Sigma^r_{\lambda} - (\Sigma_{\lambda}^{r})^{\dagger}]$, which, as will see, appears in the conductance formula. These matrices depend on the lead  spectral functions $\Gamma_{\lambda}(\omega) = \hat{v}A_{\lambda}(\omega)\hat{v}^{\dagger}$. This formula can be exploited to calculate the lesser self-energies, which are defined in the same manner, namely $\Sigma^<_{\lambda}(\omega) = \hat{v}g^<_{0,\lambda}\hat{v}^{\dagger}$. They can be rewritten as
\begin{equation}
\label{SM_eq:Sigmalesser}
    \Sigma^{<}_{\lambda}(\omega) = if_{\mu_{\lambda}}(\omega)\hat{v}A(\omega)\hat{v}^{\dagger} = if_{\mu_{\lambda}}(\omega)\Gamma_{\lambda}(\omega),
\end{equation}
with $f_{\mu_{\lambda}}$ the Fermi-Dirac distribution  corresponding to the chemical potential $\mu_{\lambda}$.\\

With a coupling to a  cavity mode, we can prove that non-equilibrium Green functions $\tilde{G}$ satisfy the equation:
\begin{equation}
\label{SM_eq:Dyson}
    \tilde{G}(\omega) = \tilde{g}(\omega) + \tilde{g}(\omega)V\tilde{G}(\omega),
\end{equation}
where $\tilde{g},\;\tilde{G},\;V$ are block matrices:
\begin{equation}
\label{SM_eq:block}
    \begin{aligned}
        \tilde{G} &= \begin{pmatrix}
            \tilde{G}^r && \tilde{G}^< \\ 0 && \tilde{G}^a
        \end{pmatrix},\\
        V & = \begin{pmatrix}
            \hat{V} && 0 \\ 0 && \hat{V}
        \end{pmatrix}.
    \end{aligned}
\end{equation}
Note that $\hat{V}$ in the second formula of (\ref{SM_eq:block}) is written in single-body formalism, and $\tilde{g}$ is defined in the same way as $\tilde{G}$. Since we only consider the correlators within the system $S$, equation (\ref{SM_eq:Dyson}) could be re-written as  $\tilde{G} = \tilde{g} + \tilde{g}V\tilde{g} + \tilde{g}V\tilde{g}V\tilde{G}$. After matrix multiplication, for the system $S$ we get: 
\begin{equation}
\label{SM_eq:Dyson2}
    \begin{aligned}
        G_S^r &= g_S^r + g_S^r\Sigma^r_{\text{leads}}G_S^r,\\
        G_S^< &= G^r_S\left( \Sigma^<_{\text{leads}} +  [g^r_S]^{-1}g^<_S[g^a_S]^{-1}\right)G^a,
    \end{aligned}
\end{equation}
where the self-energies $\Sigma^{r,<}_{\text{leads}}$ are defined above. The first expression of (\ref{SM_eq:Dyson2}) can be rewritten as $[G^r_S]^{-1} = [g^r_S]^{-1} - \Sigma^r_{\text{leads}} = [g_{0S}^r]^{-1} - \Sigma^r_{\text{int}} - \Sigma^r_{\text{leads}}$. The second formula involving lesser Green function can be proven by the relation $g^<_S = g^r_S\Sigma_{\text{int}}^<g^a_S$. 
\begin{figure}[b]
\includegraphics[width = 0.96\hsize]{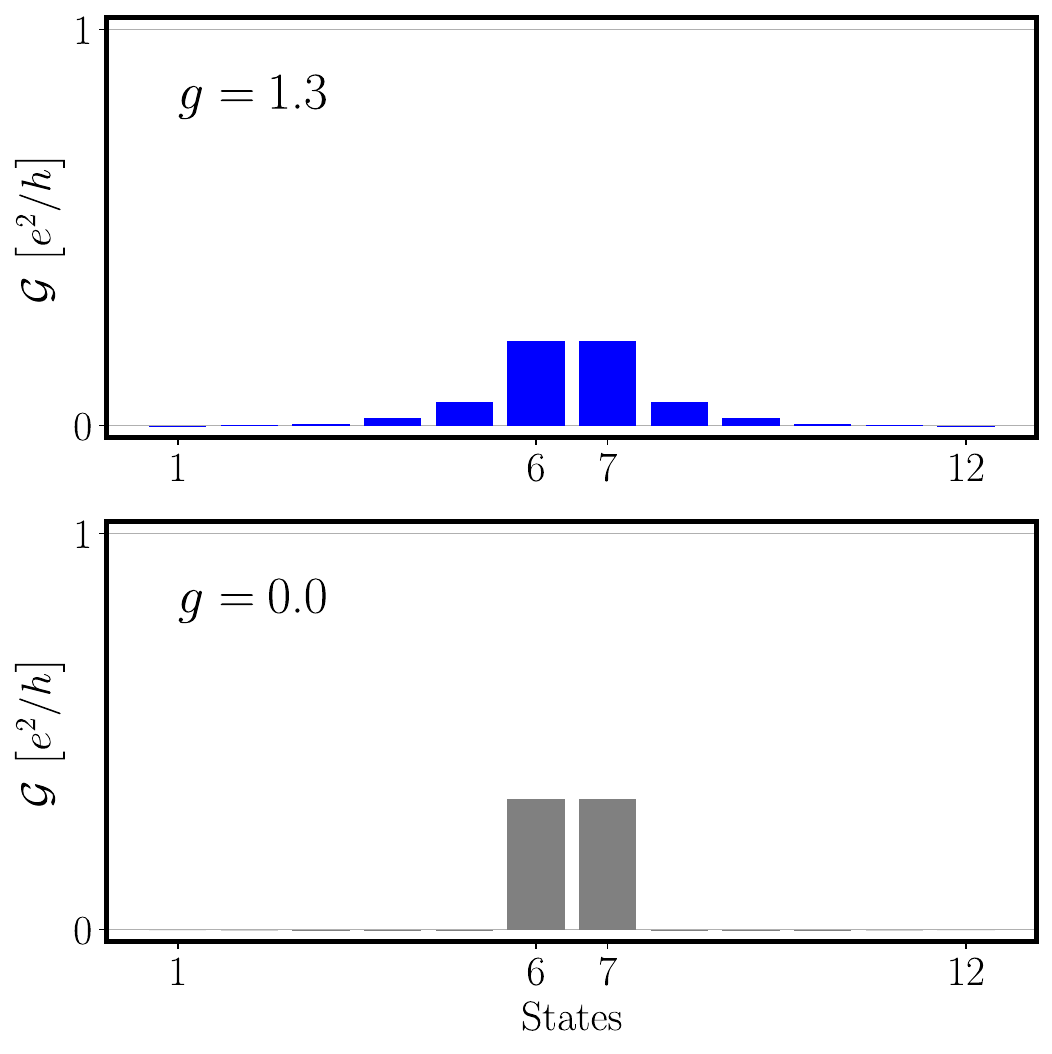}
  \caption{Contribution of each of each bare SSH states to the non-quantized conductance peaks $\mu = 0$ in Fig. 4 of the manuscript. Top (bottom) panel corresponds to the top (bottom) case in the same figure.}
\label{SM_fig:Contri}
\end{figure}
\section{Linear conductance at zero temperature}
From this point forward, we refer only to the system's Green functions, so we omit the $S$ in the expressions in order to simplify the notation.
\label{SM:Conductance}
 Meir and Wingreen derived a Landauer formula for a strongly interacting region, namely 
 \begin{equation}
\label{SM_eq:J}
    \begin{aligned}
        J = \frac{ie}{2h}\int d\epsilon &\text{Tr}\{\left[f_L(\epsilon)\Gamma^L - f_R(\epsilon)\Gamma^R\right]\left(G^r - G^a\right)\}\\ 
        + &\text{Tr}\{\left(\Gamma^L - \Gamma^R\right)G^<\},
    \end{aligned}
\end{equation}
where $J$ is the current flowing through the system, and $f_{L,R}(\epsilon) = f(\mu \pm V/2, \epsilon)$ is the Fermi-Dirac distributions with chemical potential $\mu \pm V/2$ for the left and right lead, respectively. $G^{r,a,<}$ are the system's Green functions, accounting for both leads and light-matter interaction. They can be dealt using the Keldysh formalism. In the following, we will restrict to the linear response regime (i.e, $V \rightarrow 0$) and in the limit of zero temperature. In the case of \textit{dynamical} decoupling between system and leads, we precisely compute $\Sigma_{\text{int}}^r(\omega,N_e)$ from (\ref{SM_eq:int-Sigmar}) by minimizing $\hat{\mathcal{H}} - \mu \hat{\mathcal{N}}$, where $\hat{\mathcal{N}}$ is the fermionic number operator. In the limit of weak tunneling between leads and system, we assume that $\Sigma_{\text{int}}^{r,<}$ is unaffected. In the linear conductance regime we have the following relations:
\begin{equation}
    \begin{aligned}
        f_{L,R}(\epsilon) &= f(\mu,\epsilon) \pm \frac{V}{2}\frac{\partial f}{\partial \mu}(\mu,\epsilon),\\
        \Sigma^<_{\text{leads}}(\epsilon) &= if(\mu,\epsilon)\left[\Gamma_L(\epsilon) + \Gamma_R(\epsilon)\right]\\
        &+ i\frac{V}{2}\frac{\partial f}{\partial \mu}(\mu,\epsilon)\left[\Gamma_L(\epsilon) - \Gamma_R(\epsilon)\right].
    \end{aligned}    
\end{equation}
At zero temperature, $\partial f /\partial \mu(\mu,\epsilon) = -\delta(\epsilon - \mu)$. The conductance is calculated as $\mathcal{G} = \text{lim}_{V \rightarrow 0}\left[J(V) - J(0)\right]/V$, resulting in:
\begin{equation}
    \label{SM_eq:G}
    \mathcal{G} = \frac{e^2}{h}\text{Tr}\left[\frac{i}{4}\Gamma^p\left(G^r -G^a\right)-\frac{1}{4}\Gamma^m G^r \Gamma^m G^a\right].
\end{equation}
In the above equation, we define $\Gamma^{p,m} = \Gamma^{L} \pm \Gamma^{R}$ and omit the dependence on the chemical potential for clarity.

Eq. (\ref{SM_eq:G}) can be expressed as $\mathcal{G} = \sum_{\lambda}\mathcal{G}_{\lambda\lambda}$, where $\lambda$ represents a generic basis in which the single-body Green's function is defined. By choosing these as the eigenstates of the bare SSH Hamiltonian (without the light-matter interaction), the contribution of each SSH state to the overall conductance can be determined. For the non-quantized $\mathcal{G}$s in Fig. 4 of the main text, the contributions of each state are depicted in Fig. \ref{SM_fig:Contri}. Since we consider an SSH chain with 6 unit cells, the topological edge states correspond to state $6$ and $7$. Indeed, Fig. \ref{SM_fig:Contri} clearly shows that, both with (bottom) and without (top) light-matter interaction, the total conductance for finite-size chains is dominated by the edge states.

\section{Results with a spatially-varying cavity mode field}
Our theoretical framework can be also applied to the situation of a cavity mode whose spatial profile is not flat. Here, we consider as illustrative example a cavity spatial profile described by a vector potential with a constant gradient, namely $A_0 \rightarrow A_0 + \alpha (r - L/2)$, where $L = (N-1)a_0 + b_0$ is the total length of the SSH chain, with $N$ being the number of unit cells, $a_0$ the lattice constant, and $b_0$ the intra-cell distance. This leads to a space-dependent Peierls phase $e^{-ig_v(\hat{a}+\hat{a}^{\dagger})}\hat{c}^{\dagger}_{n,B}\hat{c}_{n,A}$ and $e^{-g_w(\hat{a}+\hat{a}^{\dagger})}\hat{c}^{\dagger}_{n+1,A}\hat{c}_{n,B}$, where: 
\begin{equation}
\label{Reply1-eq:gvw}
    \begin{aligned}
        g_v \rightarrow g_v(n) &= gu + \delta g u \left(n - \frac{N+1}{2}\right),\\
        g_w \rightarrow g_w(n) &= g\left(1 - u\right) + \delta g \left(1 - u\right)\left(n - \frac{N}{2}\right), 
    \end{aligned}
\end{equation} 
with $\delta g = e\alpha a_0^2/\hbar$. Without the spatial dependence, light-matter interaction transitions the system from a trivial to a topological phase, resulting in the combination and eventual disappearance of conductance peaks. \begin{figure}[b!]
\includegraphics[width = 0.95\hsize]{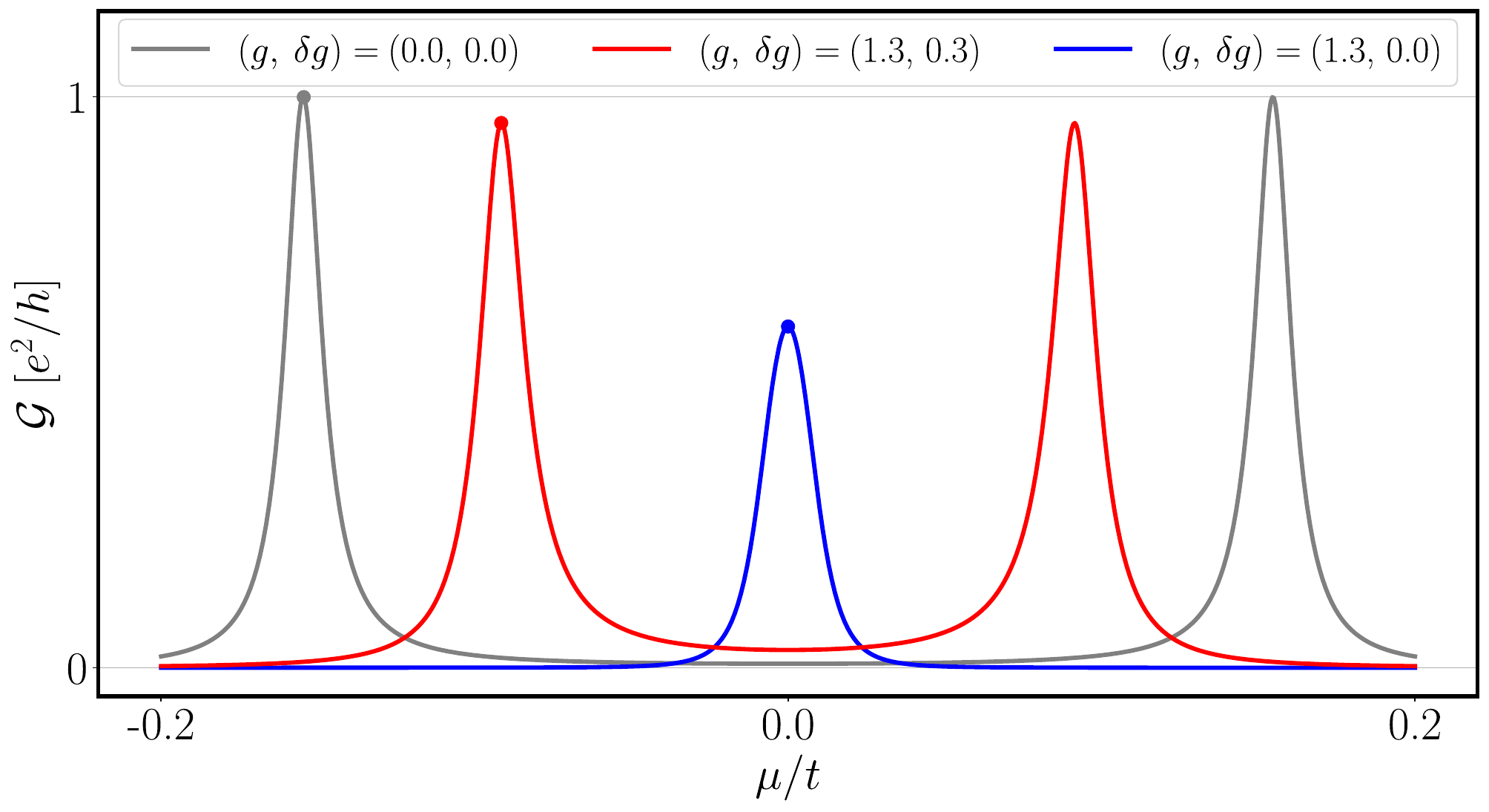}\\
\includegraphics[width = 0.95\hsize]{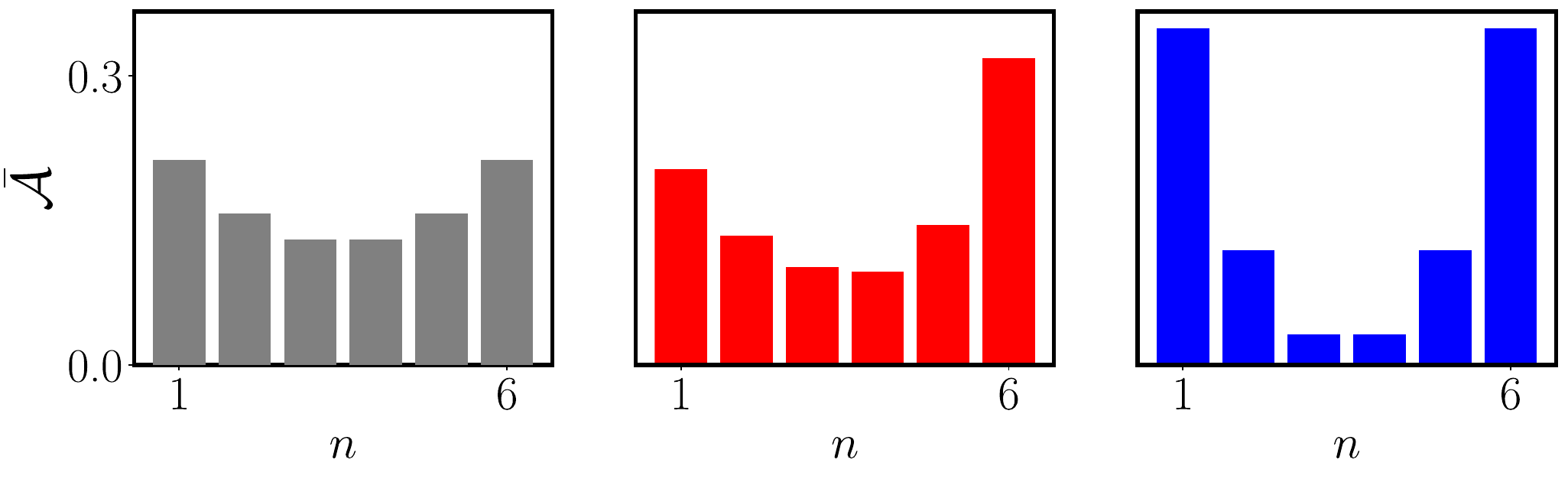}
  \caption{Top panel: Electron conductance with different coupling strengths due to the homogeneous and linear vector potential $g$ and $\delta g$. The parameters are the same as in the top panel of Fig. 4 in the manuscript. Bottom panel: Renormalized spatial spectral function calculated at the conductance peaks shown in the top panel.}
\label{SM_fig:inhomogeneous}
\end{figure} 
In this scenario of a spatially-varying cavity mode field, the conductance peaks lose their quantization. This can be explained by the fact that Eq. (\ref{Reply1-eq:gvw}) breaks inversion symmetry. The bottom panel of Fig. \ref{SM_fig:inhomogeneous} presents the electronic local spatial-dependent spectral function $\bar{\mathcal{A}}_{n}(E) = - 2\sum_{\sigma}\text{Im}\:G^r_{n,\sigma}(E)$ measured at the conductance peaks, where $G^r_{n,E}(E)$ is the retarded Green's function and $E$ the excitation energy. Here we have considered the energy of the lowest many-body excited state. The calculations clearly show that with a finite $\delta g$, $\bar{\mathcal{A}_{n}}$ loses inversion symmetry, leading to different coupling to the two leads. Due to this asymmetry, the conductance peaks are no longer quantized. These characteristics are also observed when the cavity fields induce the SSH chain to transition from a topological to a trivial phase.
\onecolumngrid

\section{Mean-field theory}
Here we provide details about the mean-field theory obtained by assuming the separable ansatz $\vert GS^{(e-p)} \rangle = \vert \psi^{(e)} \rangle \vert \chi^{(p)} \rangle$ for the electron-photon ground state. This approximation (no light-matter entanglement) yields two effective Hamiltonians, namely $\hat{\mathcal{H}}_e^{\text{eff}} = \langle \chi^{(p)} \vert \hat{\mathcal{H}}_S \vert \chi^{(p)} \rangle$ for the electronic system and $\hat{\mathcal{H}}_p^{\text{eff}} = \langle \psi^{(e)} \vert \hat{\mathcal{H}}_S \vert \psi^{(e)} \rangle$ for the photonic one. To obtain $\vert \psi^{(e)} \rangle$ and $\vert \chi^{(p)} \rangle$, we solve in a self-consistent manner the ground states for the two Hamiltonians. The two effective Hamiltonians read:
\begin{equation}
    \begin{aligned}
        \hat{\mathcal{H}}_e^{\text{eff}} &= \hbar\omega_c\langle \chi^{(p)} \vert \hat{a}^{\dagger}\hat{a}\vert \chi^{(p)} \rangle + \left(v\langle \chi^{(p)} \vert e^{-ig_v\left(\hat{a}+\hat{a}^{\dagger}\right)}\vert \chi^{(p)} \rangle\sum_{n=1}^{N}\hat{c}^{\dagger}_{n,B}\hat{c}_{n,A} + w\langle \chi^{(p)} \vert e^{+ig_w\left(\hat{a}+\hat{a}^{\dagger}\right)}\vert \chi^{(p)} \rangle\sum_{n=1}^{N-1}\hat{c}^{\dagger}_{n+1,A}\hat{c}_{n,B} + \text{h.c}\right), \\
        \hat{\mathcal{H}}_p^{\text{eff}} &= \hbar\omega_c\hat{a}^{\dagger}\hat{a}  + \left(ve^{-ig_v\left(\hat{a}+\hat{a}^{\dagger}\right)}\sum_{n=1}^{N}\langle \psi^{(e)}\vert\hat{c}^{\dagger}_{n,B}\hat{c}_{n,A}\vert \psi^{(e)}\rangle + w e^{-ig_w\left(\hat{a}+\hat{a}^{\dagger}\right)}\sum_{n=1}^{N-1}\langle \psi^{(e)}\vert\hat{c}^{\dagger}_{n+1,A}\hat{c}_{n,B}\vert \psi^{(e)}\rangle + \text{h.c}\right).
    \end{aligned}
\end{equation}
\onecolumngrid

Note that in the mean-field approximation the impact on the electronic part is essentially a renormalization of the parameters $v$ and $w$.  Indeed, we have the renormalized parameters $\tilde{v}$ = $v\langle \chi^{(p)} \vert e^{-ig_v\left(\hat{a}+\hat{a}^{\dagger}\right)}\vert \chi^{(p)} \rangle$ and  $\tilde{w} = w\langle \chi^{(p)} \vert e^{+ig_w\left(\hat{a}+\hat{a}^{\dagger}\right)}\vert \chi^{(p)} \rangle$.
\onecolumngrid

\section{Additional results with disorder}
\onecolumngrid

\begin{figure}[h!]
\includegraphics[width = 0.45\hsize]{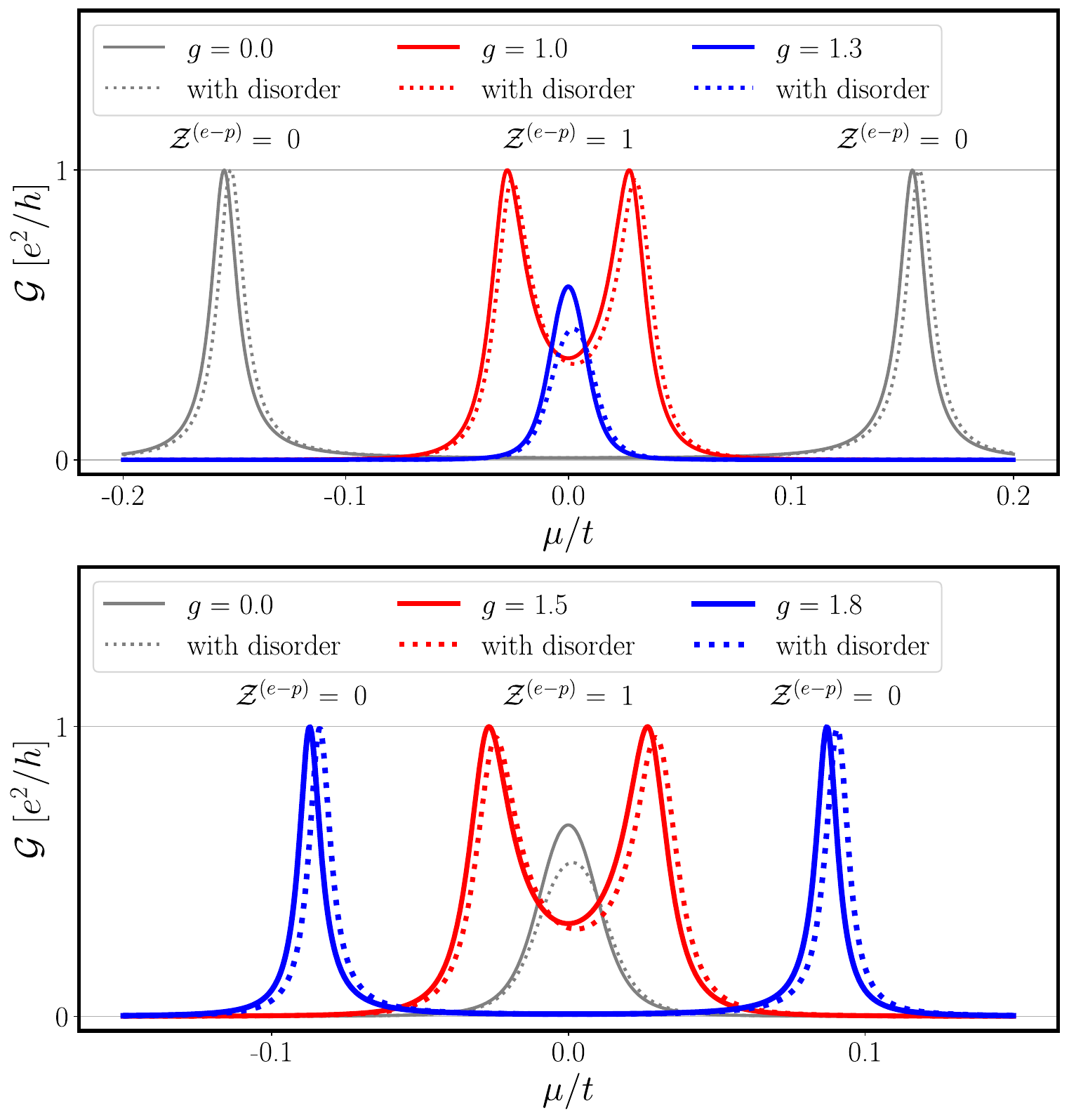}
\includegraphics[width = 0.45\hsize]{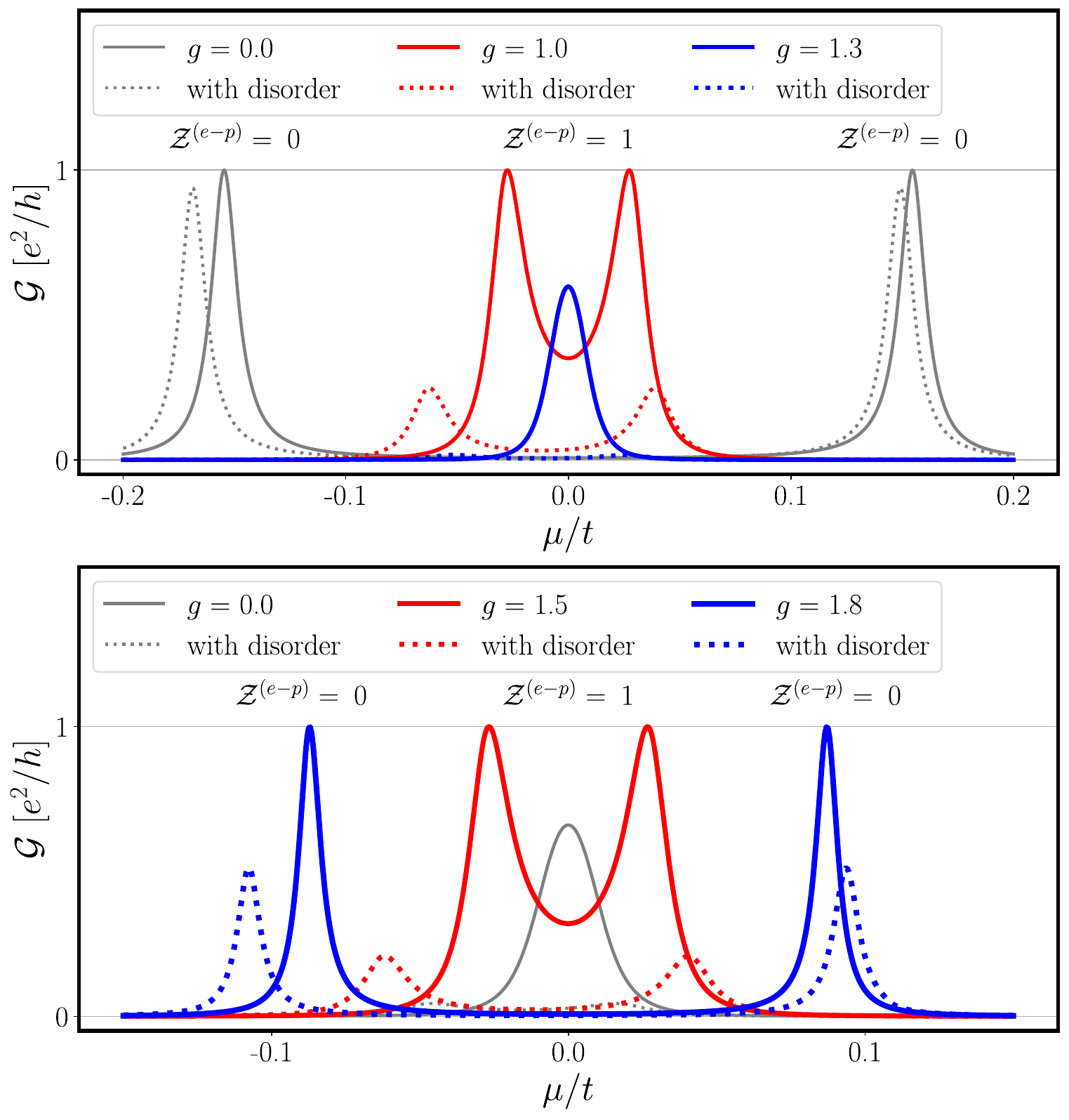}
\caption{Same conductance curves as in Fig. 4 of the main manuscript, but with the disorder amplitude $W = 0.015t$ (left panels) and $W = 0.15t$ (right panels).}
\label{SM_fig:W}
\end{figure}
Here we show conductance curves as in Fig. 4 of the main manuscript, but with different values of the disorder amplitude $W$, namely $W = 0.015t$ and $W = 0.15t$
for Fig. \ref{SM_fig:W}.
\twocolumngrid
\bibliography{Bib.bib}

\end{document}